\documentclass[aps,prl,twocolumn,showpacs,superscriptaddress,10pt]{revtex4-1}
\usepackage{graphicx} 
\usepackage{amssymb} 
\usepackage{amsmath}
\usepackage{mathrsfs}
\usepackage{soul,xcolor}

\begin{document}

\setstcolor{blue}

\title{THz generation from relativistic plasmas driven by near- to far-infrared laser pulses}

\author{J. D{\'e}chard}
\author{X. Davoine}
\author{L. Berg\'e}
\email{luc.berge@cea.fr}
\affiliation{CEA, DAM, DIF, F-91297 Arpajon, France}

\date{\today}

\begin{abstract}
Terahertz pulse generation by ultra-intense two-color laser fields ionizing
gases with near- to far-infrared carrier wavelength is studied from
particle-in-cell (PIC) simulations. For long wavelength (10.6 $\mu$m) promoting
a large ratio of electron density over critical, photoionization is shown to
catastrophically enhance the plasma wakefield, causing a net downshift in the
optical spectrum and exciting THz fields with tens of GV/m
amplitude in the laser direction. This emission is accompanied by coherent
transition radiation (CTR) of comparable amplitude due to wakefield-driven
electron acceleration. We analytically evaluate the fraction of
CTR energy up to $30\,\%$ of the total radiated emission including the particle
self-field and numerically calibrate the efficiency of the
matched blowout regime for electron densities varied over three orders of
magnitude.
\end{abstract}

\maketitle

The ability of terahertz (THz) waves to probe matter is attracting interest
for lots of applications reviewed, e.g., in
\cite{Berge:epl:126:24001}. Recently, novel challenges such as
compact THz electron
accelerators \cite{Nanni:nc:6:8486,Sharma:jpb:51:204001,Curcio:sr:8:1052} or
THz-triggered chemistry \cite{LaRue:prl:115:036103} raised the need of mJ THz
pulses with high field strength $>$ GV/m. Optical rectification in organic
crystals \cite{Vicario:prl:112:213901} or using tilted-pulse-front pumping
\cite{Fulop:oe:22:20155} can achieve percent conversion efficiency and sub-mJ
THz energies with few 0.1 GV/m  field strengths. These solid-based technologies,
however, remain limited by damage thresholds. In contrast, gas plasmas created
by intense, two-color laser pulses may supply
suitable emitters free of any damage \cite{Hafez:jo:18:093004}. Electrons are
tunnel-ionized by the asymmetric light field usually composed
of a near-IR fundamental wavelength (800 nm) and its second harmonic (400 nm).
Their ``photocurrent'' polarized in the laser direction generates a broadband
photocurrent-induced radiation (PIR) in the THz range \cite{Kim:np:2:605}.
Nevertheless, two-color setups using moderate intensities $\sim 10^{14}$
W/cm$^2$ only achieve conversion efficiencies $< 10^{-3}$ and
$\mu$J energies \cite{Oh:apl:102:201113,Meng:apl:109:131105}.

In the relativistic regime, however, when the normalized
vector potential $a_0 \equiv 8.5 \times 10^{-10} I_{0}^{1/2} [\mbox{W/cm}^2]
\lambda_0 [\mu\mbox{m}]$ is larger than unity ($I_{0}$ is the intensity and
$\lambda_0$ denotes the laser wavelength), plasma waves trigger a strong
longitudinal field exploited for laser-wakefield acceleration (LWFA)
\cite{Tajima:prl:43:267}. Accelerated electrons crossing the plasma-vacuum
interface can then emit coherent transition radiation (CTR)
operating in the THz band. Leemans et al.
\cite{Leemans:prl:91:074802,Leemans:pop:11:2899} reported THz energy of 3-5 nJ
per pulse measured from a dense gas jet of helium. Theoretical estimates
assuming a Boltzmann distribution of 4.6 MeV confirmed this energy yield and
anticipated  the possibility to provide few 100 $\mu$J energies by increasing
the electron bunch to tens of MeV and/or the plasma diameter to mm scales.
Record values were later achieved in numerical simulations from which CTR took
over PIR by delivering conversion efficiencies~$> 5\times 10^{-3}$ and mJ THz
pulse energies \cite{Dechard:prl:120:144801}. Such performances have been
reached in laser-solid experiments \cite{Gopal:ol:38:4705,Liao:pnas:5:3994}.

Besides, mid- and far-infrared light sources supplying TW peak powers are today
available. Femtosecond laser facilities with $3.9\,\mu$m central wavelength
opened the way to multi-octave supercontinuum generation
\cite{Kartashov:ol:37:3456} and can accelerate electrons to 12 MeV energy in gas
jets \cite{Woodbury:ol:43:1131}. CO$_2$ lasers ($\lambda_0 = 10.6\,\mu$m) are
also operational in the ps range \cite{Tochitsky:np:13:41} and they
should soon provide revolutionary tools unveiling new
regimes in particle acceleration and future colliders
\cite{Pogorelsky:prab:19:091001}. Therefore, it is worth investigating the gain
that such optical sources may offer in THz science, since their carrier
wavelength is already close to the spectroscopy range of interest. Several
studies \cite{Clerici:prl:110:253901,Nguyen:pra:97:063839,
Fedorov:oe:26:31150,Tulsky:pra:98:053415} highlighted the impressive growth in
the THz energy yield when increasing the pump wavelength of two-color-gas setups
at moderate intensities. Relativistic interactions remain to be
explored for this purpose.

The content of this Letter is threefold. First, along the laser polarization
axis, we report from PIC simulations a drastic change in the laser-to-THz
conversion process when long-wavelength pulses interact with gases in
relativistic regime. The underlying mechanism differs from both PIR and CTR:
Here, ionization fronts increase the ponderomotive pressure
and catastrophically enhance the plasma waves, which, in the self-modulated (SM)
LWFA regime, downshift the laser spectrum to low frequencies $<$ 10 THz.
Second, in the radial direction, better electron injection
gives rise to a strong CTR field with comparable amplitude $\sim\,20$ GV/m.
Third, we quantify the radiated energy for electron densities
over critical covering three decades and provide an estimate of the CTR yield
disconnected from the particle self-field.

Simulations are performed with the PIC, kinetic code \textsc{calder}
\cite{Lefebvre:nf:43:629,Gonzalez:sr:6:26743} solving Maxwell-Vlasov equations
with strong-field ionization \cite{Ammosov:spjetp:64:1191} in 2D planar
geometry. The longitudinal (transverse) axis is set along $x$ (resp. $y$), while
the laser pulse is linearly polarized in the $z$ direction. We here consider
three fundamental wavelengths: $\lambda_0 \equiv 2\pi c / \omega_0 = 0.8$, $3.9$
and $10.6$ $\mu$m for the same laser potential $a_0=2.2$, associated to the
input intensities $I_0 \simeq 10.5,\,0.44$ and $0.06 \times 10^{18}$ W/cm$^2$,
respectively. The input pulse is Gaussian in time and space with its two
harmonics ($\omega_0,\, 2\omega_0$) having the same FWHM duration $\tau_0=150$
fs and FWHM transverse width $w_0$ varying between 20 and
$50\ \mu$m. The intensity ratio between the two harmonics is 10 $\%$ and their
initial phase shift is $\pi/2$. The two-color field is focused into a gas cell
of helium with atomic density $n_a=5.5 \times 10^{17}$ cm$^{-3}$, along a
trapezoidal density profile of length $L_p$ with $200\ \mu$m-long plateau and
$25\ \mu$m ramps. The plasma wavelength is $\lambda_p = \sqrt{n_c/n_e}
\lambda_0$, where $n_c \simeq 1.11\times 10^{21}/\lambda_0^{2}[\mu\mbox{m}]$ is
the critical density and $n_e$ denotes the electron density.
The ratio $n_e/n_c$ in He thus increases from $6.3 \times
10^{-4}$ to 0.11 for $0.8 \leq \lambda_0 \leq 10.6\,\mu$m while $\lambda_p =
32\,\mu$m. The frequency window chosen to extract the THz waveforms is $\nu
\equiv \omega/2\pi < \nu_0/3$, which is performed by inverse
Fourier Transform using a 6th-order hyperGaussian filter.

\begin{figure}
      \includegraphics[width=\columnwidth]{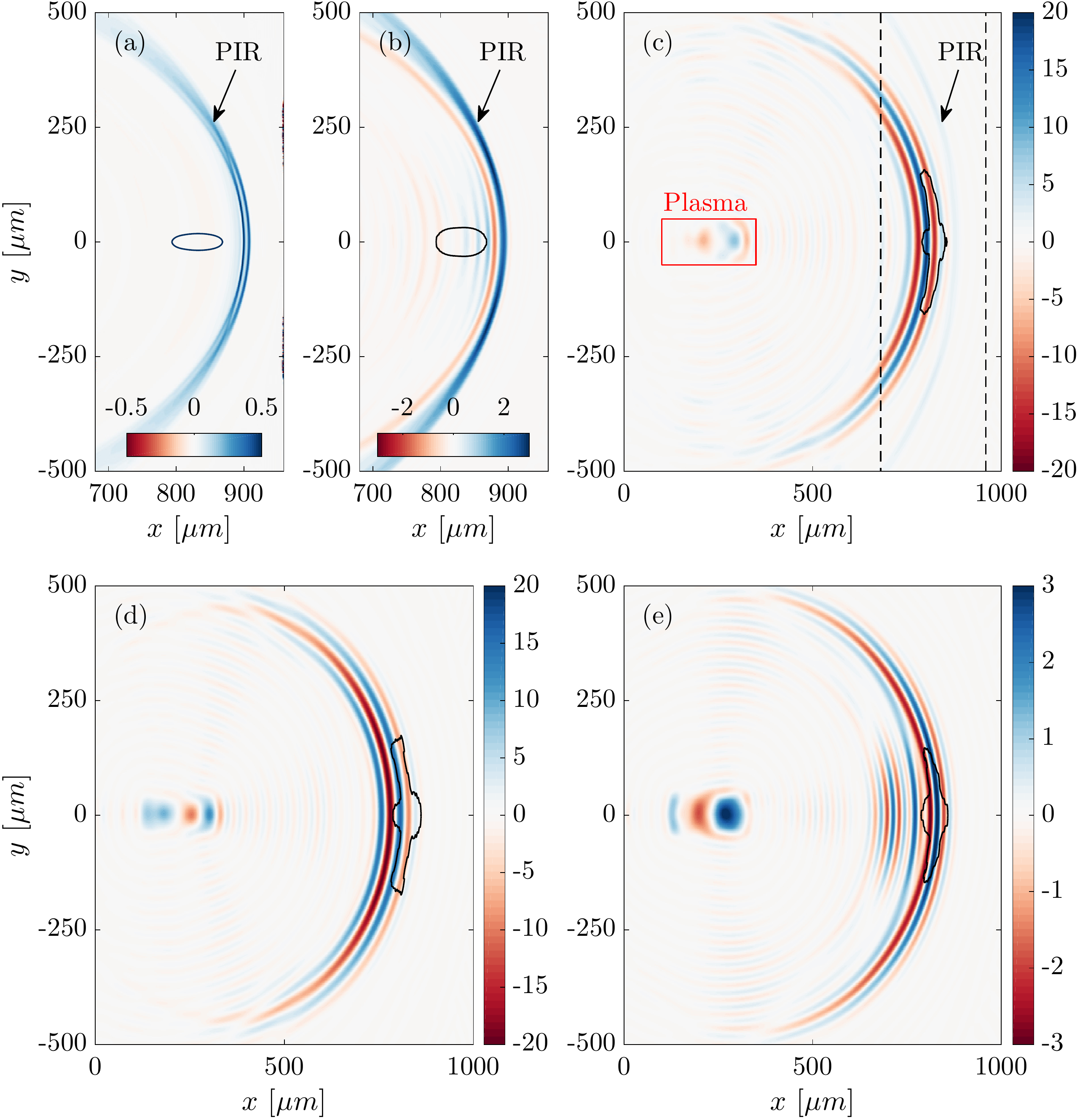}
      \caption{(a-c) Electric field $E_z$ (color bars in GV/m) produced by a
two-color Gaussian pulse with $a_0 = 2.2$ ionizing He, filtered in the range
$\nu < \nu_0/3$ and transmitted to vacuum at $t=3200$ fs ($500\ \mu$m after the
plasma channel, see \cite{SupMat}) for $\lambda_0=$ (a) 0.8, (b) 3.9 and (c)
10.6 $\mu$m. Arrow points to the PIR field. In (c) the black dashed rectangle
indicates the simulation domain of (a,b) encompassing the laser region; the red
rectangle delineates the plasma volume. (d,e) show the same field pattern for
(d) a single-color pulse and (e) a pre-ionized plasma at 10.6 $\mu$m. Gray
contours are eye-guides for the laser pulse envelope undergoing distortions
along propagation.}
      \label{Fig1}
\end{figure}

{\it Laser-polarized THz field} - Figure \ref{Fig1}(a) shows a snapshot of the
$z$-polarized THz field for $\lambda_0=0.8\,\mu$m at 500 $\mu$m after the
plasma-vacuum interface. Two PIR peaks emerge with about 0.5 GV/m maximum
amplitude. When increasing the laser wavelength to $\lambda_0 = 3.9$ $\mu$m, the
PIR becomes more efficient since the electron transverse momentum, $p_z$, and
related current density linearly scale with $\lambda_0$
\cite{Debayle:oe:22:13691}, which is confirmed by the field strength (3 GV/m) of
Fig. \ref{Fig1}(b). With $\lambda_0=10.6\ \mu$m, however, photocurrents only
deliver the first wavefront on the right-hand side of Fig. \ref{Fig1}(c), being
much weaker than expected ($\sim 2$ GV/m, see arrow). This sudden drop of PIR is
attributed to the fall in the photocurrent efficiency, decreased as the
ionization sequences develop together with the wakefield. Surprisingly, a
multi-cycle THz pulse localized behind the PIR with 20~GV/m field strength and
$\sim 7$ THz oscillation frequency emerges, which seems triggered by another
process as its location, emission angle and amplitude are
different. To tackle this THz waveform, other simulations have been performed
with one color only, yielding a similar pattern without the PIR field [Fig.
\ref{Fig1}(d)]. Furthermore, when simulating a pre-ionized plasma, the THz field
strength decreases by a factor $\sim 6$ [compare color bars of Figs.
\ref{Fig1}(c) and \ref{Fig1}(e)]. Hence, for long-wavelength pumps,
photoionization keeps a non-trivial impact on the transverse THz emission, but
its action differs from the standard PIR mechanism. 

\begin{figure}
     \includegraphics[width=\columnwidth]{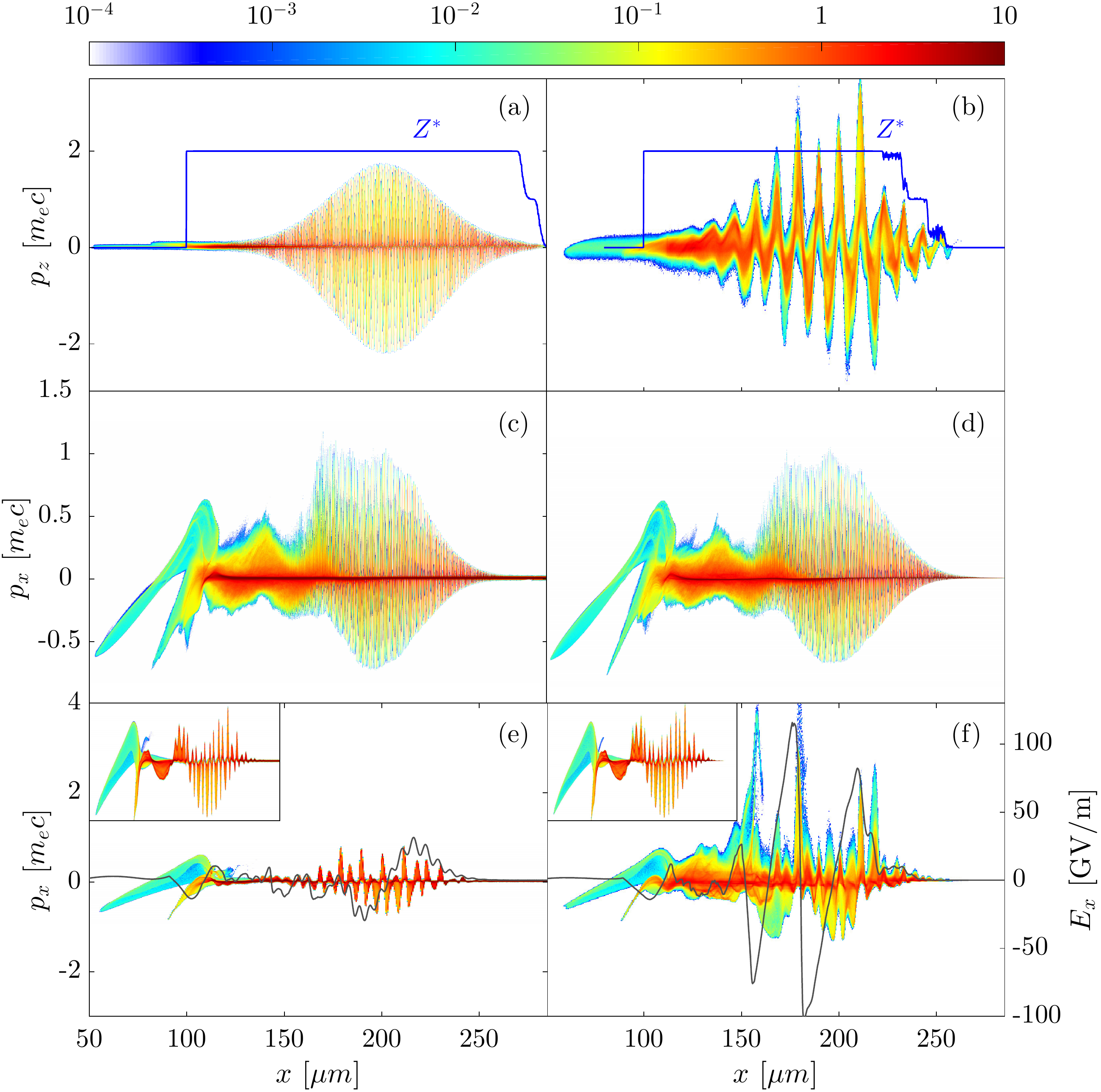}
      \caption{(a,b) Transverse momentum at $t=1100$ fs for $\lambda_0=$ (a) 0.8
and (b) 10.6 $\mu$m. The blue curves show the ionization degree of He. (c,d)
Longitudinal momentum for $\lambda_0=0.8\ \mu$m in (c) pre-ionized helium and
(d) helium undergoing ionization. (e,f) same information for $\lambda_0=10.6\
\mu$m. Gray curves display the longitudinal electric field (right axis). Insets
show the same phase space $(x,p_x)$ for hydrogen $(n_e = n_a)$.}
      \label{Fig2}
\end{figure}

Proving the importance of photoionization, Figs. \ref{Fig2}(a,b) display the
$(x, p_z)$ electron phase space when the laser is fully inside the plasma for
$\lambda_0=0.8$ and 10.6 $\mu$m, respectively. The blue curves plot the growth
in the ion charge $Z^*$ along the optical path. At each ionization instant,
freed electrons acquire a kick in their transverse momentum $p_z$, linked to the
laser vector potential \cite{Mori:prl:69:3495}. For $\lambda_0 = 0.8\,\mu$m, the
transverse drift momentum initiated by ionization and exiting the rear pulse is
small. In contrast, for the $10.6\,\mu$m pulse, $p_z$ reaches higher values in
the ionization zone. Figures \ref{Fig2}(c-f) compare the longitudinal phase
space $(x, p_x)$ at the same time in pre-ionized or initially-neutral helium.
The longitudinal momentum develops two characteristic oscillations, i.e., the
$\sim\,2\omega_0$ fast component of the laser ponderomotive force and the plasma
frequency $\sim \omega_p$. The plasma wave modulates the pulse envelope in the
SM-LFWA regime since $c\tau_0 = 45\ \mu$m $> \lambda_p$. No noticeable change
occurs at 0.8 $\mu$m, whether or not ionization is acting, due to weak momentum
transfer. However, compared with pre-ionized He, $p_x$ with $\lambda_0 = 10.6$
$\mu$m reaches much higher values, up to 5 $m_ec$ [Fig. \ref{Fig2}(f)], and thus
increases the electrostatic field that develops a sawtooth-like nonlinear
plasma wave. Photoionization hence directly impacts the wakefield dynamics,
in particular by increasing the ponderomotive pressure exerted
on the plasma by particles born at rest \cite{Mori:prl:69:3495,SupMat}. Insets
in Figs. \ref{Fig2}(e,f), showing no change for a hydrogen gas, indeed evidence
that amplification of the plasma waves in He requires secondary ionization
events after the formation of the wakefield. These properties are justified in
\cite{SupMat} by means of a 1D, quasi-static model accounting for ionization in
the transverse and longitudinal momenta of the plasma wave.

\begin{figure}
      \includegraphics[width=\columnwidth]{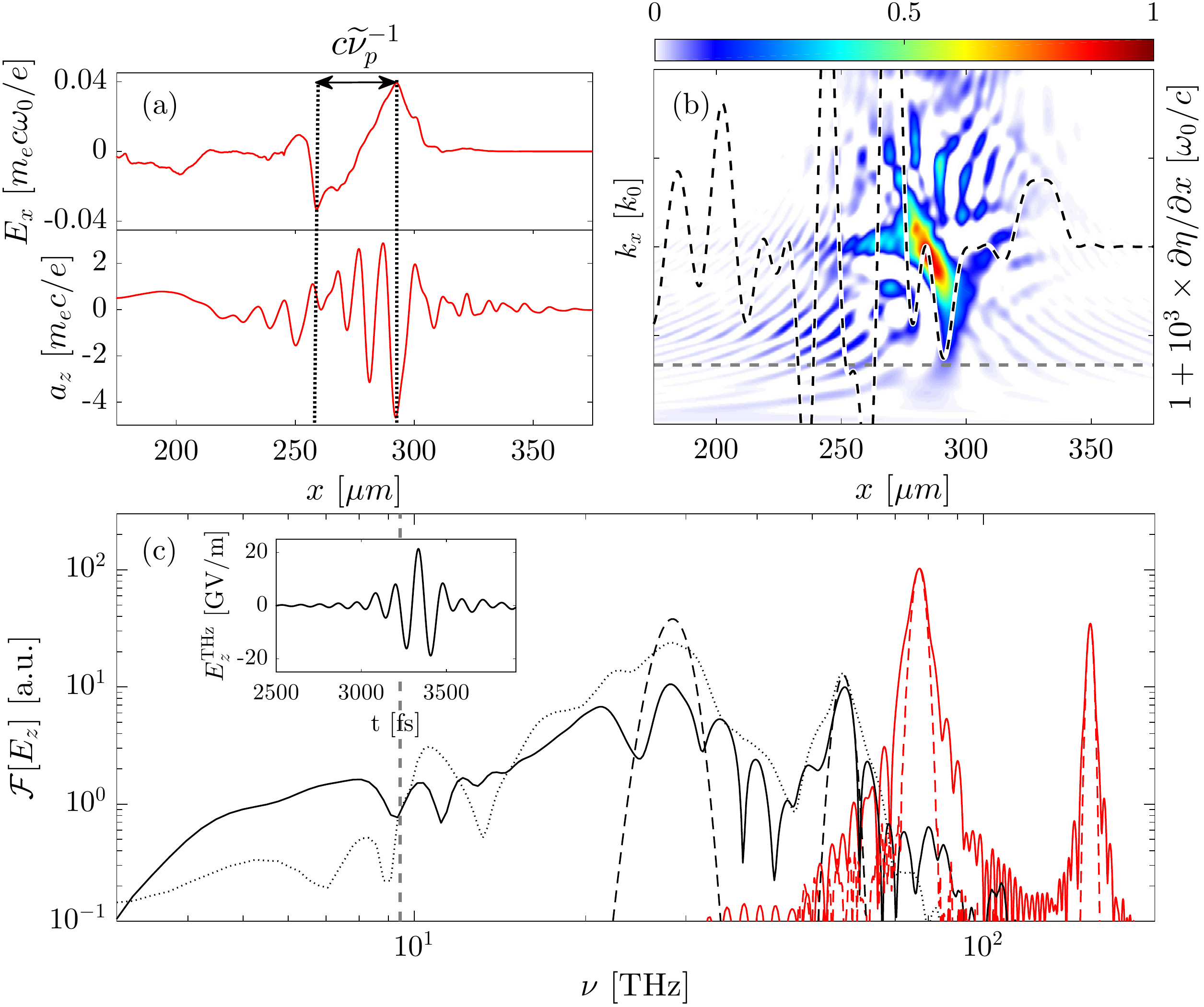}
      \caption{(a) On-axis PIC longitudinal field (top) and transverse vector potential (bottom) at $t=1400$ fs. (b) Wigner Transform $\mathcal{W}_E(x,k) = \int_{-\infty}^{+\infty}E(x+x'/2)E^*(x-x'/2)e^{-ikx'}\, dx'$ of the on-axis two-color laser electric field for $\lambda_0 = 10.6\,\mu$m along with the gradient of the refractive index computed from the ratio $n_e/n_c \gamma$ using PIC simulation data (dotted black curves). The dashed line indicates the cut-off frequency ($9.33$ THz) and wavenumber. (c) On-axis log-log scaled spectrum of $E_z$ at the entrance of the simulation domain (dashed black curves) and at $x=840\,\mu$m (solid curves) for ionized He. Inset: THz field. Red curves: $3.9\,\mu$m, black curves: $10.6\,\mu$m. The dotted black curve shows the on-axis spectrum for 10.6 $\mu$m in preionized helium.}
      \label{Fig3}
      \end{figure}

\begin{figure}
      \includegraphics[width=\columnwidth]{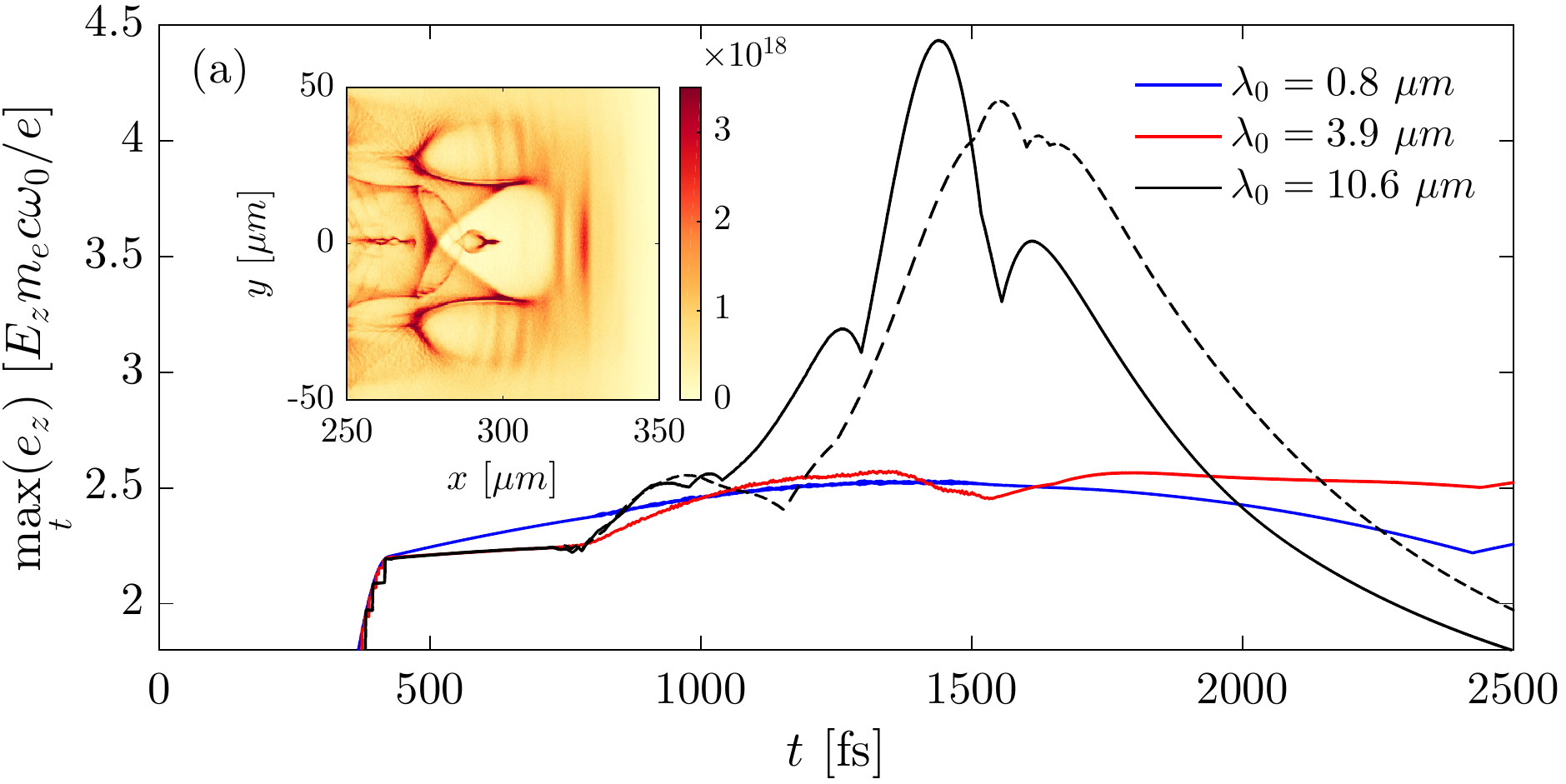}
      \includegraphics[width=\columnwidth]{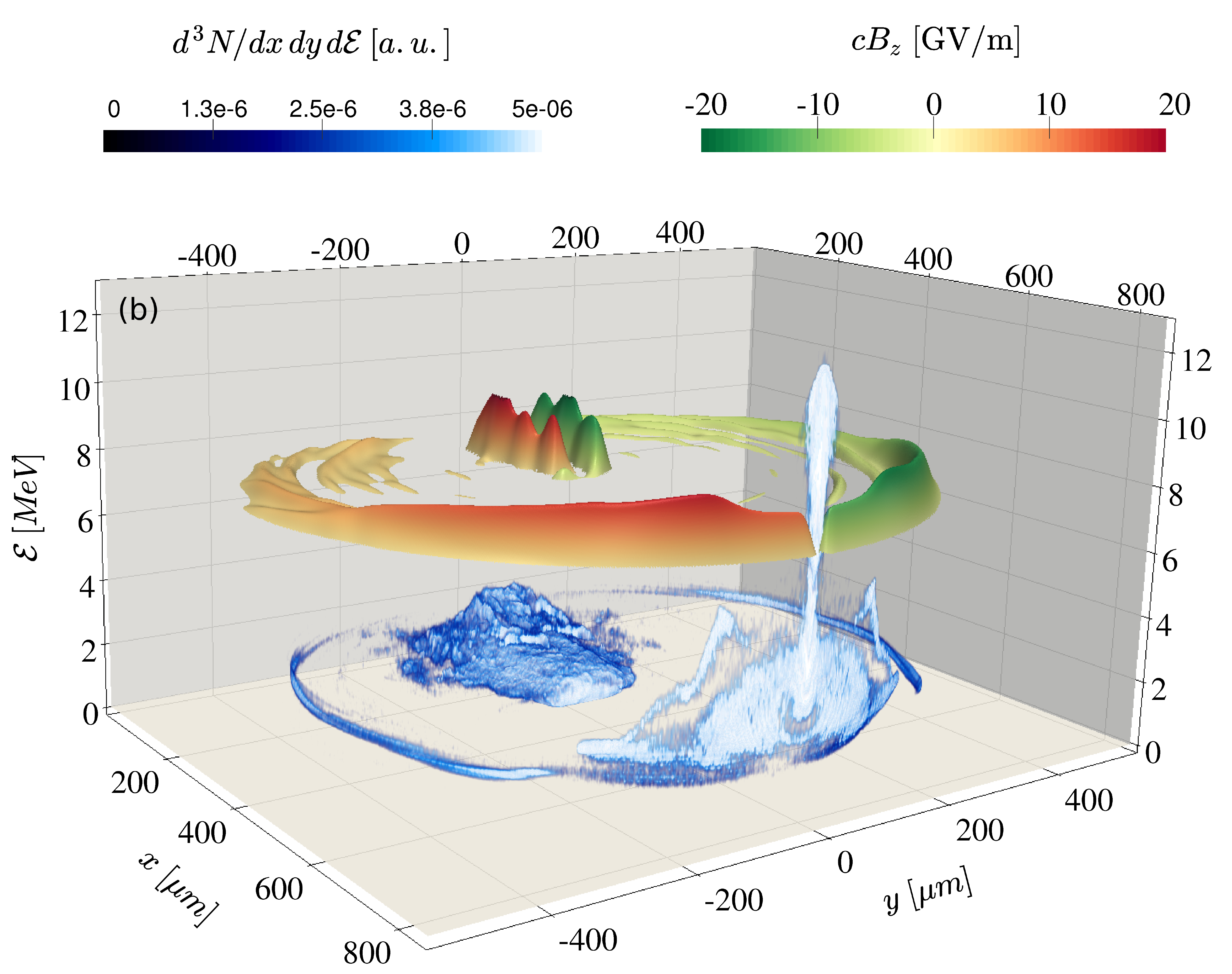}
      \caption{(a) Maximum normalized two-color laser field amplitude for the three wavelengths $\lambda_0$. The dashed curve shows $e_z$ for pre-ionized helium when $\lambda_0=10.6\,\mu$m. Inset displays a plasma bubble where electron injection takes place at the plasma exit ($t=1400$ fs). (b) Density map (blue colormap) in the phase space $(x,y,\mathcal{E})$ at $t = 2500$ fs, where $\mathcal{E}$ [MeV] denotes the electron energy. On top of it the radiated CTR magnetic field $c B_z$ [GV/m] is plotted (green-red colormap).}
      \label{Fig4}
\end{figure}

Figure \ref{Fig3}(a) illustrates the PIC on-axis longitudinal and transverse
field in ionized He. The THz field is centered near the relativistic plasma
frequency ${\tilde \nu}_p \equiv \nu_{p}/\sqrt{\gamma} \sim 7$ THz of the
wakefield modulating the rear pulse, where $\gamma$ denotes the
electron Lorentz factor. The feedback of the sharp plasma wave fluctuations on
the laser pulse nonlinearly modifies the optical refractive index, $\eta \approx
\sqrt{1-n_e/n_c\gamma}$, which promotes the creation of new wavelengths
\cite{Mori:ieee:33:1942}. The ponderomotive force accumulates electrons, which
decreases the pulse group velocity (photon deceleration) and leads to local
redshifts. Figure \ref{Fig3}(b) shows the Wigner transform of the on-axis laser
electric field near the exit of the plasma channel. We clearly observe a
frequency downshift ($\partial_x k < 0$) at the top of the plasma out-ramp
induced by the propagation in the plasma along which the
gradients in the refractive index, $\eta \partial_x \eta = - \frac{1}{2}
\partial_x n_e/n_c \gamma$ $\propto \partial_x k$, are also
detailed. In the frequency domain, Fig. \ref{Fig3}(c) shows the log-log
amplitude spectrum of the transverse field transmitted to vacuum for the mid-IR
fundamental wavelengths investigated. Unlike the 3.9 $\mu$m pump, the 10.6
$\mu$m laser spectrum (black curves) widely broadens between 1 and 100 THz
around the pump wave and develops a net enlargement around the plasma frequency.
With a ratio $n_e/n_c$ equals to $0.11$, Raman instabilities modulate the laser
pulse prior to wakefield acceleration in the bubble regime
\cite{Gordon:pre:64:046404}. The latter is attained when the pulse width is
close to $(\lambda_p/\pi) \sqrt{a_z}$ \cite{Lu:prstab:10:061301} and high-field
values $a_z > 4$ are reached by self-focusing [see Fig. \ref{Fig4}(a)]. The
optical spectrum then slips down to less than 8 THz. Figure
\ref{Fig3}(c) evidences that the photon redshift around ${\tilde \nu}_{p} \simeq
7$ THz is particularly marked when ionization is activated.
This mechanism explains the THz field patterns of Fig.
\ref{Fig1}(c) and we evaluate its efficiency to $1.2\%$ in the frequency window
$\nu < 9.5$ THz.

{\it Radially-polarized THz field} - High amplitudes in the
longitudinal field [Fig. \ref{Fig2}(f)] promote particle injection. Using longer
laser wavelengths also favors self-focusing for peak powers above critical, $P >
P_c = 17 (\lambda_p/\lambda_0)^2$[GW], and increases the charge of the
accelerated electron bunch \cite{Woodbury:ol:43:1131}. Figure \ref{Fig4}(a)
shows the maximum normalized laser electric field along the propagation axis for
the three studied wavelengths. The 0.8 and 3.9-$\mu$m pumps do not self-focus
due to the weak ratio $n_e/n_c$. In contrast, a clear sequence of collapse and
plasma blowout occurs at $\lambda_0=10.6\ \mu$m, where the intensity is
amplified by a factor $\sim 4$ and reinforced by pulse steepening through
multi-ionization of He. Photoionization thus accelerates the bubble formation.
As a result, the laser spot enters the blowout wakefield regime
\cite{Lu:prstab:10:061301} leading to electron injection into a well-shaped
bubble [see inset of Fig. \ref{Fig4}(a)]. Later, this electron bunch will pass
the plasma-vacuum boundary, generating a CTR field radiating in the THz range
\cite{Leemans:prl:91:074802,Dechard:prl:120:144801}. To visualize the
correlation between the accelerated particles and the CTR field, Fig.
\ref{Fig4}(b) shows a map revealing three distinct populations of electrons. The
first one constituting the plasma channel has rather low energy and is located
near axis at $y \approx 0,\,x \leq 400\,\mu$m (see blue area). The second
electron population forms the expanded bubble outside the
plasma at $y \approx \pm 400\,\mu$m. At $x \simeq 700\,\mu$m, the third,
wakefield-accelerated electron population reaches an energy as high as 12 MeV.
The CTR field generated by the escaping electrons is represented in 2D by the
$z$-polarized magnetic field $B_z$ $= \int (\partial_y E_x -
\partial_x E_y) dt $ (green-red colormap). This single-cycle field has a
maximum amplitude of about 20 GV/m and corresponds to $1.4\%$ conversion
efficiency ($\nu < 9.5$ THz). Associated to 3D laser pulses with
222 mJ input energy (focal spot radius equal to $w_0$), this THz field may thus
contain $\sim 3$ mJ energy. Such CTR pulses,
radially-polarized in 3D, could be distinguished from the former laser-polarized
THz component by using a THz analyzer handling ellipticity
\cite{Kosareva:ol:43:90}.

Finally, we parametrically scan the impact of the density ratio $n_e/n_c$,
previously varied through $\lambda_0$, on the transition radiation. Because CTR
is mainly driven by the fundamental pump, we simulate a single-color 1-$\mu$m,
35-fs FWHM laser pulse with 3.7 J energy using the \textsc{calder-circ} code
\cite{Dechard:prl:120:144801}. The density ratio is increased from $n_e/n_c
\simeq 4.5 \times 10^{-4}$ up to unity. Accordingly,
$\lambda_p$ decreases from 47 $\mu$m down to 1 $\mu$m, while the laser pulse
resonantly excites the wakefield whenever $\lambda_p \sim 10\,\mu$m, i.e.,
$n_e/n_c \sim 10^{-2}$. Our simulations are constrained by a constant areal
density, $n_e L_p = 7.8 \times 10^{17}$ cm$^{-2}$, for treating the same
electron number in the longitudinal direction. To keep a unique electron bunch,
the transverse beam waist may be decreased when the highest density ratios lead
to multiple filamentation ($P/P_c > 10$).

\begin{figure}
      \includegraphics[width=0.9\columnwidth]{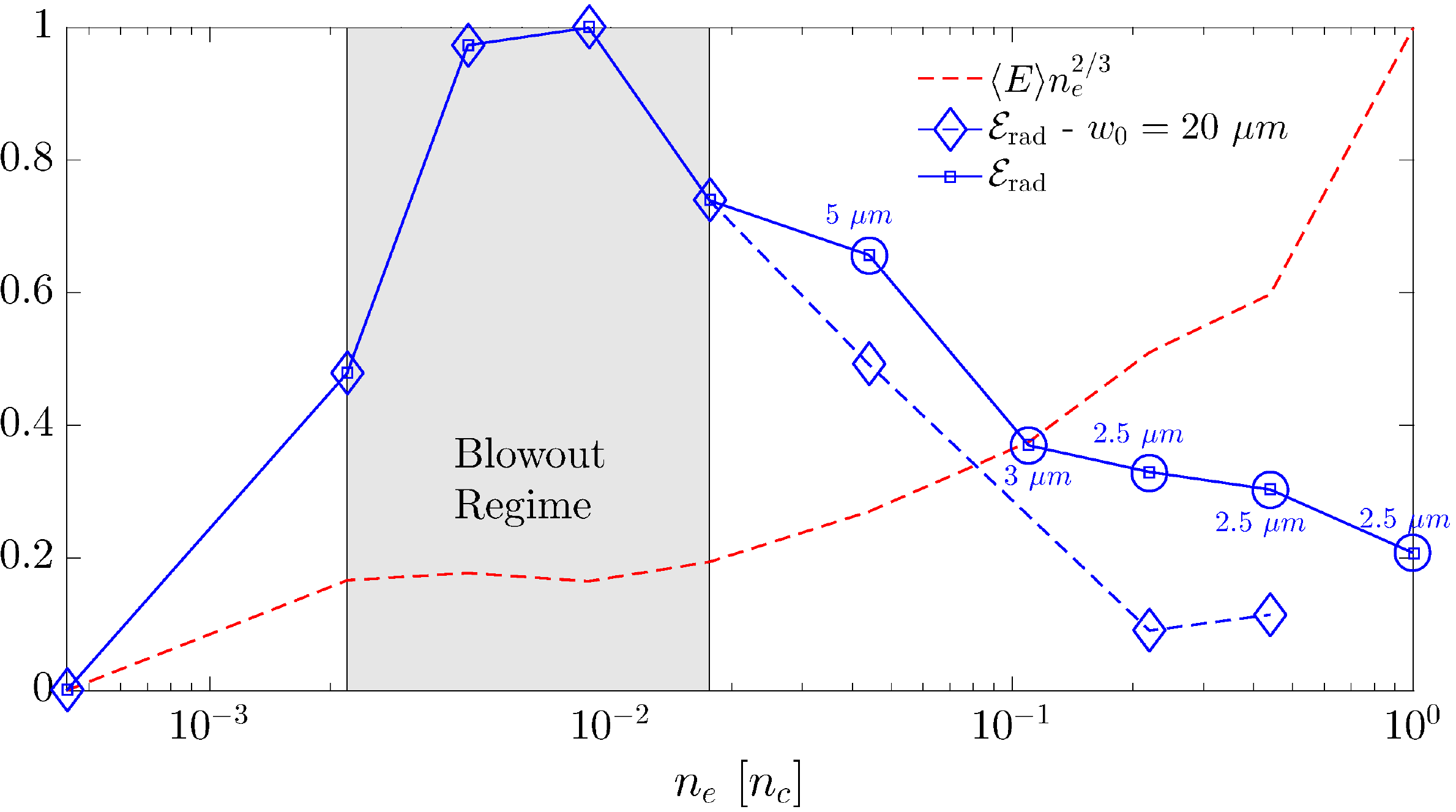}
      \caption{THz energy in the range $\nu < 0.3\,\nu_0$ (blue solid lines)
normalized to its maximum value (39 mJ) for 1-$\mu$m single-color pump at $x =
500$ $\mu$m from the vacuum-plasma interface and $\langle E \rangle n_e^{2/3}$
(red dashed line) extracted from PIC simulations as function of the electron
density ([$n_c$]). Blue circles specify values of the laser waist decreased from
$20\,\mu$m to avoid filamentation. Gray area delineates the blow-out regime.}
      \label{Fig5}
\end{figure}

Figure \ref{Fig5} summarizes the radiated energy computed from ten plasma
densities (blue curve). The dashed red curve plots variations in the product of
the mean electron energy $\langle E \rangle$ and $n_e^{2/3}$, expected to be
constant in the matched blowout regime \cite{Lu:prstab:10:061301}. Maximum
radiated field is reached in this regime and field emission by
wakefield-accelerated electrons only varies by a factor $\sim 5$ over three
plasma decades. For $n_e/n_c > 10^{-2}$, electron acceleration enters the
SM-LWFA regime implying a drop in the energy yield, which is reinforced by
multiple filamentation occuring with large waist ($w_0 = 20\,\mu$m, dashed blue
curve). Inspection of data reveals the generation of up to 50 GV/m electric
fields (not shown).

These impressive field values are, however, produced by both the CTR component
further reaching the detector and the self-field of the accelerated electrons.
The splitting time $t_s$ necessary for separating the CTR field propagating at
$c$ in vacuum and the self-field of an electron bunch with $\mu$m length $L_b$
is typically $t_s \approx 2 \gamma^2 L_b/c$ \cite{Carron:pier:2000}, requiring
mm length scales that are barely accessible in PIC simulations. Evaluating the
magnetic field from the Biot-Savart equation applied to a uniform bunch with
zero radius enables us to separate the CTR energy from the remaining radiated
energy. This analysis, performed in \cite{SupMat}, extracts the fraction of CTR
yield, which is found to remain within $20-30\%$ of the total radiated energy.
This result invites us to decrease all our previous CTR field amplitudes and
energies by a factor $\sim 2$ and $\sim 4$, respectively. These energy estimates
remain larger than Leemans et al.'s measurements, as the acceleration schemes
exploited in \cite{Leemans:prl:91:074802} used plasma lengths exceeding far the
electron dephasing length $\sim \gamma^2 \lambda_p$ and were out of the optimum
blowout regime.

In summary, we demonstrated that photoionization matters when long laser
wavelengths are employed to create intense THz fields in relativistic plasmas.
First, the ionization-induced pressure is non zero from the second electron
extraction, which increases the plasma wakefield and cause photon deceleration.
This dynamics generates high THz fields through frequency downshifts in the
optical spectrum. Second, CTR by the wakefield accelerated electrons is enhanced
by more efficient self-focusing. We endly evaluated the fraction of CTR energy
in gases up to $\sim$ 30$\%$ of the overall radiation emitted in the near-field
and demonstrated that the blowout regime provides the highest
wakefield-accelerated energy yield. A laser-to-THz conversion efficiency close
to the percent could be reached, demonstrating another relevance of CO$_2$
lasers in relativistic laser-plasma interaction.

\section{Acknowledgement}
The authors acknowledge L. Gremillet, G. Sary and A. Debayle for fruitfull discussions as well as GENCI France for awarding us access to the supercomputer CURIE using Grant $\#$ A0010506129. 

\bibliography{references_prl}
\end{document}